\documentclass[aps, twocolumn, superscriptaddress]{revtex4-2} 

\usepackage[switch]{lineno}
\usepackage{amsmath}
\usepackage{empheq}
\usepackage[parfill]{parskip}
\usepackage[active]{srcltx}
\usepackage{color}
\usepackage{array}
\usepackage{booktabs}
\usepackage{amsfonts}
\usepackage{dsfont}
\usepackage{graphicx}
\usepackage{hyperref}

\begin{document}

\setlength{\parindent}{0.5cm}

\title{The forced one-dimensional swarmalator model}

\author{Md Sayeed Anwar}
\email{sayeedanwar447@gmail.com}
\affiliation{Physics and Applied Mathematics Unit, Indian Statistical Institute, 203 B. T. Road, Kolkata 700108, India}

\author{Dibakar Ghosh}
\email{diba.ghosh@gmail.com}
\affiliation{Physics and Applied Mathematics Unit, Indian Statistical Institute, 203 B. T. Road, Kolkata 700108, India}

\author{Kevin O'Keeffe}
\email{kevin.p.okeeffe@gmail.com}
\affiliation{Senseable City Lab, Massachusetts Institute of Technology, Cambridge, MA 02139}  

\begin{abstract}

%\hspace{1 cm}  (Received XXXX; accepted XXXX; published XXXX) \\

We study a simple model of swarmalators subject to periodic forcing and confined to move around a one-dimensional ring. This is a toy model for physical systems with a mix of sync, swarming, and forcing such as colloidal micromotors. We find several emergent macrostates and characterize the phase boundaries between them analytically. The most novel state is a swarmalator chimera, where the population splits into two sync dots, which enclose a `train' of swarmalators that run around a peanut-shaped loop. 

\noindent
%DOI: XXXXXXX
\end{abstract}

\maketitle

%%%%%%%%%%%%%%%%%%%%%%%%%%%%%%%%%%%%%
\section{Introduction}
Synchronization is a common type of pattern formation that occurs in many systems, from fireflies \cite{buck1988synchronous} to lasers \cite{kozyreff2000global} to heart cells \cite{peskin75}. In the synchronized state, the units coordinate the timing of their oscillations, but do not move through space. Swarming is a complementary form of self-organization where the units coordinate their motion but do not have internal phase variables that self-synchronize. Examples are the flocking of birds \cite{bialek2012statistical}, schooling of fish \cite{katz2011inferring}, and the herding of sheep \cite{garcimartin2015}. 

Though related -- in sense spatiotemporal opposites -- sync and swarming have historically developed independently. New research has however brought the two fields into contact by considering oscillators that are mobile, their movements and phase oscillations mutually coupled. These entities are called swarmalators, since they generalize swarms and oscillators. Swarmalators are expected to be useful as toy models for the many systems which combine sync and self-assembly such as chemical motors and biological microswimmers \cite{tamm1975role,verberck2022wavy,belovs2017synchronized,yan2012linking,hwang2020cooperative,zhang2020reconfigurable,bricard2015emergent,manna2021chemical,riedel2005self,yang2008cooperation}.

Recently, researchers have started to map out the space of swarmalator phenomenology by studying physics-style, minimal models. Simple mean field models, where all swarmalators' interact with equal strength, produces sync disks and vortex-like phase waves seen in arrays of sperm and active colloids \cite{o2017oscillators}. Adding pinning led to chaos, quasi-periodicity, and other unsteady behavior that matched the dynamics of magnetic domain walls \cite{sar2024solvable,sar2023pinning}. Time delays led to pseudo-crystalline states with slow dynamics reminiscent of glasses  \cite{blum2024swarmalators}. Local coupling \cite{lee2021collective}, thermal noise \cite{hong2018active}, mixed-sign interactions \cite{mclennan2020emergent,jimenez2020oscillatory}, and other effects  \cite{lizarraga2023synchronization,ghosh2023antiphase,yadav2024exotic,kongni2023phase,anwar2024collective,o2018ring,smith2024swarmalators,ansarinasab2024spatial,ansarinasab2024spatial,schilcher2021swarmalators,degond2022topological,lizarraga2024order,gong2024approximating} have also been explored.

Yet one effect has yet to be understood: external forcing. Forcing is conspicuous in many swarmalator systems such as colloidal micromotors \cite{yan2012linking,yan2013colloidal} but is difficult to analyze theoretically due to the nonlinearities and numerous degrees of freedom at play. A first step in this direction was carried out a few years ago \cite{lizarraga2020synchronization}. The authors took the two-dimensional (2d) swarmalator model \cite{o2017oscillators} and added a simple sinusoidal forcing term, and found diverse behavior. Yet even in the simpler mathematical setting of a minimal model, analysis was still intractable. 

This paper aims to advance the theoretical study of forced swarmalators by considering an even simpler model where the swarmalators' motion is confined to a one-dimensional (1d) ring. As we will show, these simplifications allow the expressions for the stabilities and bifurcations of several collective states to be derived exactly. Moreover, the model can be generalized in many ways, hopefully making it a useful tool for follow-up studies.

\begin{figure*}[t!]
	\centering
	\includegraphics[scale=0.14]{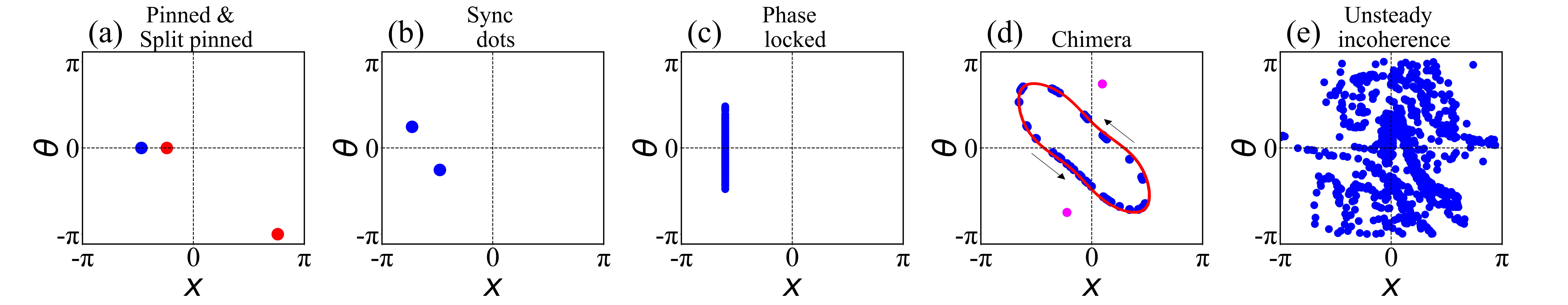}
	\caption{Collective states of the forced swarmalator model. (a) Pinned (blue dot) and split pinned (red dots) state for $(K, F)=(1,0.5)$, (b) Sync dots for $(K, F)=(-2,1)$, (c) Phase locked state for $(K, F)=(-2, 1.8)$,
    (d) Chimera state for $(K, F)=(-2,0.7)$, (e) Unsteady coherence state for $(K, F)=(-2,0.2)$. All results are found by integrating Eqs.~\eqref{eoms-x} and \eqref{eoms-theta} for $T = 1500$ for $N = 1000$ swarmalators.}
	\label{states}
\end{figure*}

%%%%%%%%%%%%%%%%%%%%%%%%%%%%%%%%%%%%%%%%%%%%%%%%%%%%%%%%%%%%%%%%%%%%%%%%%%%%%%%%%%%%%%%%%%%%%%%%%%%%%%%%%%%%%
\section{Mathematical Model}
Since our goal is to maximize tractability, we study the simplest possible model of forced swarmalators: the 1d swarmalator model \cite{o2022collective,yoon2022sync}  with sinusoidal forcing on the phase dynamics (yet as we will show, even in this radically simplified setting, the model have behaviour that remains out of analytic reach). The model is
\begin{align}
    \dot{x_i} &=  \nu + \frac{J}{N} \sum_j^N \sin(x_j - x_i) \cos(\theta_j - \theta_i), \label{eoms-x} \\
    \dot{\theta_i} &= \omega + \frac{K}{N} \sum_j^N \sin(\theta_j - \theta_i) \cos(x_j - x_i) \nonumber \\
    & \;\;\;\;\;+ F \sin(\Omega t - \theta_i). \label{eoms-theta}
\end{align}
\noindent
Here, $x_i, \theta_i \in S^1$ are the position and phase of the $i$-th swarmalator ($S^1$ is the unit circle). The spatial dynamics model aggregation (the Kuramoto $\sin(x_j-x_i)$ term) which depends on phase (the accompanying $\cos(\theta_j-\theta_i)$ factor). Conversely, the phase dynamics model synchrony which depends on distance, with the addition of the sine forcing. 

The parameters $J, K$ and $(\nu, \omega)$ are the associated couplings and natural frequencies, and $F,\Omega$ are the strength and frequency of the forcing. For simplicity, we study resonant forcing where the driving frequency is the same as the natural frequency $\Omega = \omega$. Then, by going to a suitable rotating frame, we set $\omega = \nu = 0$ without loss of generality. By rescaling time, we set $J=1$ which leaves a model in two-parameters $(K,F)$,
\begin{align}
    \dot{x_i} &= \frac{1}{N} \sum_{j=1}^{N} \sin(x_j - x_i) \cos(\theta_j - \theta_i), \\
    \dot{\theta_i} &= - F \sin \theta_i + \frac{K}{N} \sum_{j=1}^{N} \sin(\theta_j - \theta_i) \cos(x_j - x_i). 
\end{align}
The model contains a competition between the forcing $F$, which wants to pins the phases of the oscillators at $\theta = 0$ \footnote{in the co-moving frame we are in $\theta = \omega t$}, and the phase coupling $K$, which tends to maximize or minimize the oscillators phase differences for $K>0\; \text{or}\; K<0$. The overall dynamics are, however, more complex than that because the spatial dynamics also play a role. They induce spatial aggregation also and mediate the strength of the phase interactions.

\par We use the following order parameters to classify the macroscopic behavior that emerges,
\begin{align}
    & W_{\pm} = S_{\pm} e^{i \phi_{\pm}} = \langle e^{i (x \pm \theta)} \rangle, \\
    & (Z, Y) = (R e^{i \Psi_{\theta}}, Q e^{i \Psi_x}) = (\langle e^{i \theta} \rangle, \langle e^{i x} \rangle), \\
    & V = \langle \sqrt{v_{x}^2 + v_{\theta}^2} \rangle, 
\end{align}
where $\langle . \rangle$ denotes ensemble average. $W_{\pm}$ are rainbow order parameters introduced in previous studies of swarmalators \cite{o2017oscillators,o2022collective} and measure the space/phase correlation. When $x_{i} = \pm \theta_i$, they are maximal, and when $x_i$ is uncorrelated with $\theta_i$, they are minimal. $Z$ is the regular Kuramoto sync order parameter, and $Y$ a natural generalziation for the spatial degree of freedom $x$. Finally, $V$ is the mean speed of the ensemble and is included to discern static states from non-static ones.

% %%%%%%%%%%%%%%%%%%%%%%%
\section{Numerics}
\begin{figure*}[t!]
	\centering
	{\includegraphics[width=2.0\columnwidth]{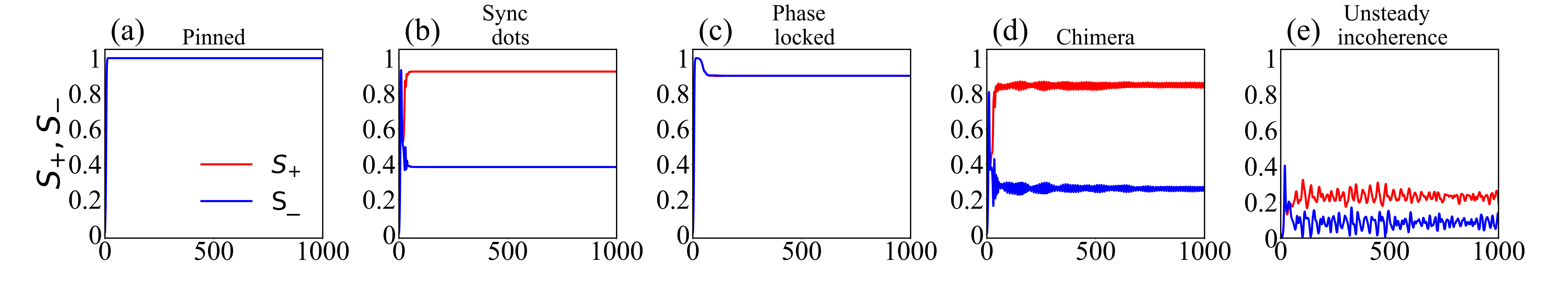}\\
		\includegraphics[width=2.0\columnwidth]{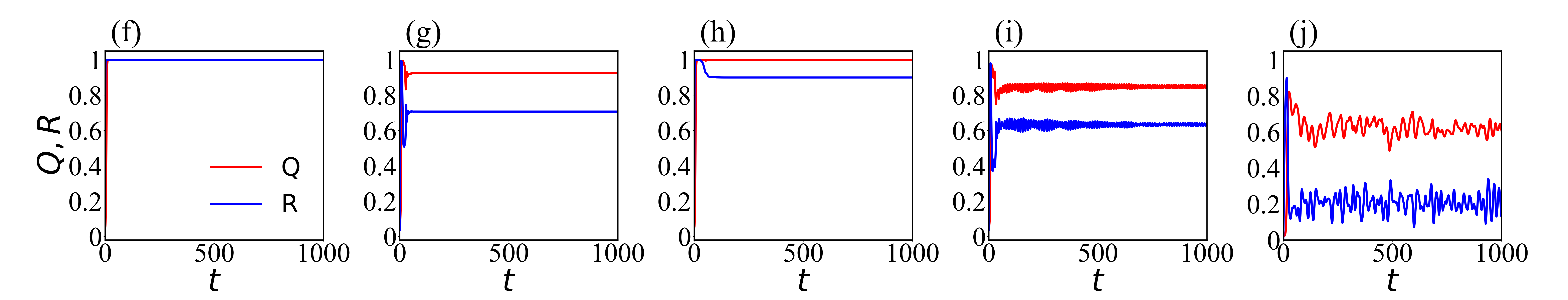}}
	\caption{(Upper panel) $S_{\pm}$ vs $t$, (lower panel) $Q, R$ vs $t$. Parameter values for each state are listed in the caption of Fig.~\ref{states}. The split pinned state is omitted for clarity. }
	\label{ops_vs_t}
\end{figure*}
Numerics show the system always reached six long-term modes of self-organization depending on the initial conditions and parameters (Figure~\ref{states}). These are

\par \textit{Pinned}. Panel (a) blue dot. Swarmalators are pinned to the driving field $\theta_i = 0$ and fully aggregated in space $x_i = C$ for constant $C$ (the constant $C$ stems from the rotational symmetry in the space equation). Figures~\ref{ops_vs_t}(a, f), show all the order parameters saturate at unity $Q = R = S_{\pm} = 1$. 

\par
\begin{figure}[t!]
	\centering
	{\includegraphics[width= 0.9\columnwidth]{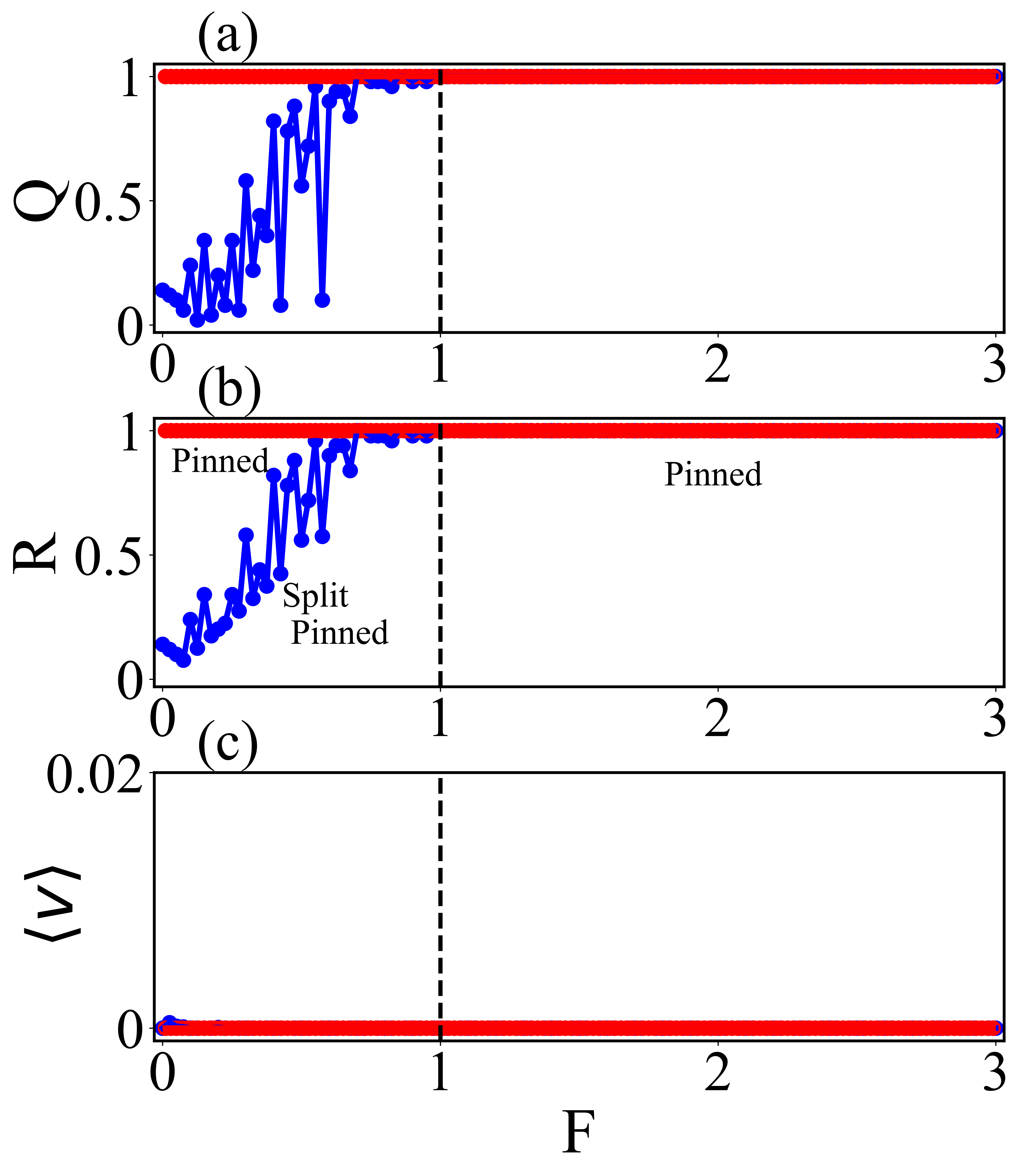}}
	\caption{Order parameters $R,Q$, and mean velocity $\langle V \rangle$ versus forcing amplitude $F$ for attractive phase coupling $K=1>0$. The pinned state (red dots) is stable for all $F$ and defined by $Q= R = 1$. The split pinned (blue dots) is stable for $F<F^* = 1$ and has $0<Q, R<1$. The vertical dashed black line is the theoretical prediction up to which the split pinned state is stable, obtained from Eq.~\eqref{split_pinned_critical}. The rainbow order parameters $S_{\pm}$ are not plotted because they do not provide any additional information. The simulation parameters used are $(N, T)=(1000,1500)$. Each data point is plotted by taking the average of the last 20\% realizations.}
	\label{R_versus_F}
\end{figure}

 \textit{Split pinned}. Panel (a) red dots (plotted on the same panel to save space). Like the pinned state above, except the swarmalators split into two camps. One at the pinning site $\theta_i =0 $, the other at $\theta_i = \pi$. The number of swarmalators in each camp depends on the initial conditions and means the sync order parameters take values between $ 0 < R, Q < 1$ (Figure~\ref{R_versus_F}).

\par \textit{Sync dots}. Panel (b) Swarmalators form two sync dots with coordinates $(x_{1},\theta_{1})=(C,\theta^{*})$, $(x_{2},\theta_{2})=(C+\Delta x,-\theta^{*})$. As before, the constant $C$ is arbitrary and stems from the rotational symmetry in the $\dot{x}$ equations. The number of swarmalators in each dot depends on the initial conditions. Two dots appear when the initial positions and phases are drawn uniformly at random from $(0,2\pi)$, and one dot when they are drawn from the $(0,\pi)$. Figures~\ref{ops_vs_t}(b, g) show $R, Q, S_{\pm} <1$.

\par
\begin{figure}[t!]
	\centering
	{\includegraphics[width= 0.9\columnwidth]{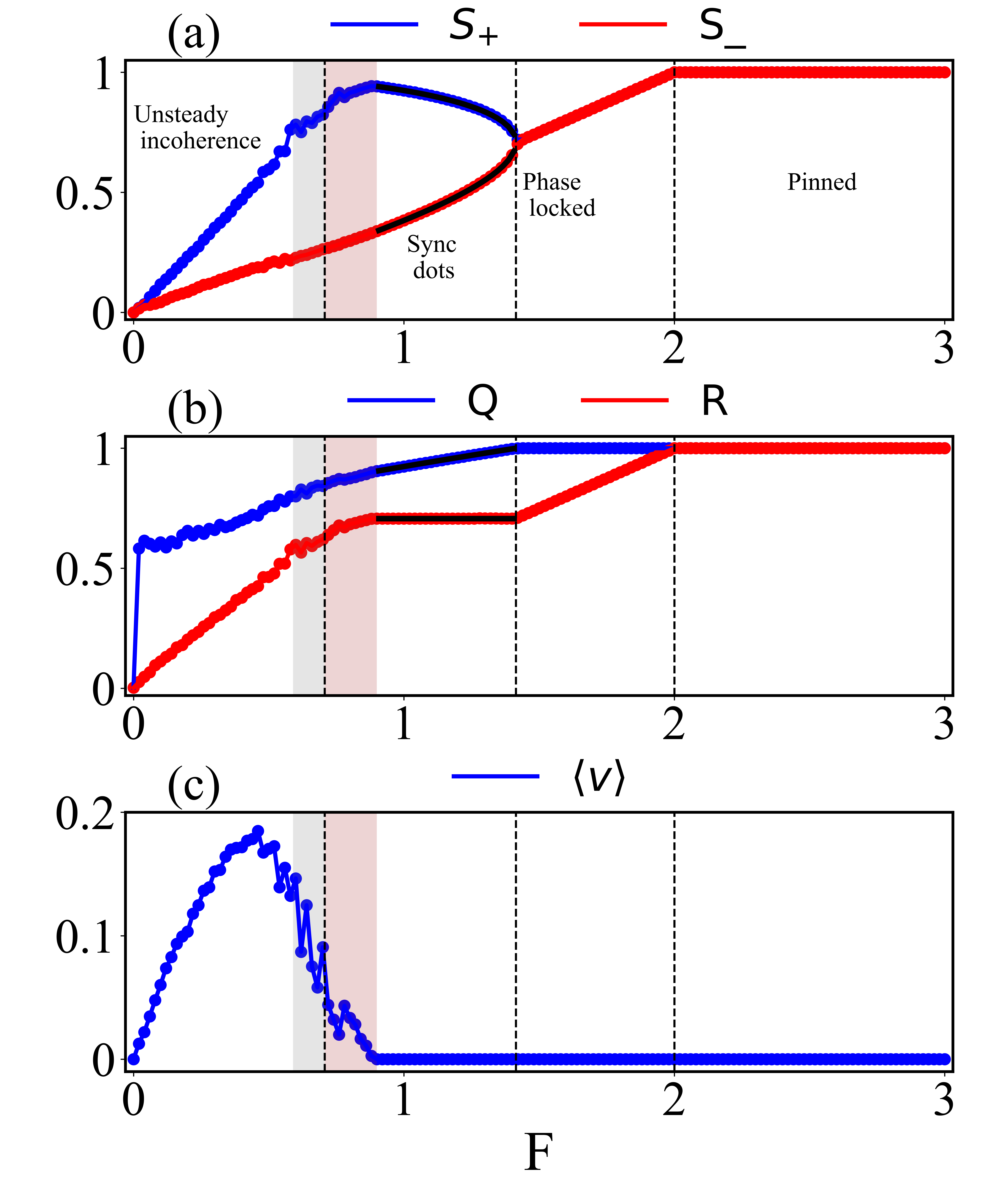}}
	\caption{Variation of order parameters :  (a) rainbow order parameters $S_{\pm}$, (b) space, phase order parameters $Q$, $R$, and (c) mean velocity $\langle V \rangle$ as a function of external forcing amplitude $F$ for repulsive phase coupling $K=-2<0$. The vertical dashed black lines are the theoretical boundaries associated with the three static states, obtained from Eqs.~\eqref{pinned_critical}, \eqref{sync_dots_critical}, and \eqref{phase_locked_critical}, respectively. The red-shaded region corresponds to the region of bistability where both `sync dots' and `chimera' states exist depending on the choice of initial condition, while the grey-shaded region represents the area where only the chimera state emerges. The simulation parameters used are $(N, T)=(1000,1500)$. Each data point is plotted by taking the average of the last 20\% realizations.}
	\label{ops_vs_F}
\end{figure}

 \textit{Phase locked}. Panel (c) Swarmalators aggregate at a single position $x_i = C_1$, but exhibit \textit{phases locking}, defined by constant phase \textit{differences} $\theta_i - \theta_j = C_{ij}$, as distinct from \textit{perfect} synchrony where $\theta_i = C_2$, constant phase. Figure~\ref{phase_locked} in the Appendix-\ref{phase_locked_histogram} shows the distribution of phases is compactly supported. Seeing phase locking, but not perfect synchrony, in a system of identical oscillators, is uncommon. For example, in the Kuramoto model with identical units, only full sync of $\theta_i$ is realized. The order parameters take values $S_{+}=S_{-}=R<1$ [Figs.~\ref{ops_vs_t}(c),~\ref{ops_vs_t}(h) and \ref{ops_vs_F}(a), \ref{ops_vs_F}(b)].

\par \textit{Swarmalator chimera}. Panel (d). This is an exotic state where the swarmalators split into three subpopulations: Two sync dots (magenta dots), which enclose a ``sync train". This train is the peanut-shaped loop colored red in the figure. This is a generalization of the chimera states in regular oscillators \cite{abrams2004chimera,kuramoto2002coexistence}. The chimera is unsteady. The sync dots oscillate back and forth slowly and with small amplitude while the swarmalators in the train execute full cycles around the loop. The movements manifest as small oscillations in the order parameters time evolution [Figs.~\ref{ops_vs_t}(d),~\ref{ops_vs_t}(i)]. You can see the dynamics most clearly in Supplementary Movie 1.

\par \textit{Unsteady incoherence}. Panel (e). The swarmalators form an incoherent and unsteady mess like a gas cloud. The state is also unsteady, with noisy vacillations in the order parameters [Figs.~\ref{ops_vs_t}(e),~\ref{ops_vs_t}(j)]. 

\begin{figure}[t!]
    \centering
    {
    \includegraphics[width=\columnwidth,height=0.65\columnwidth]{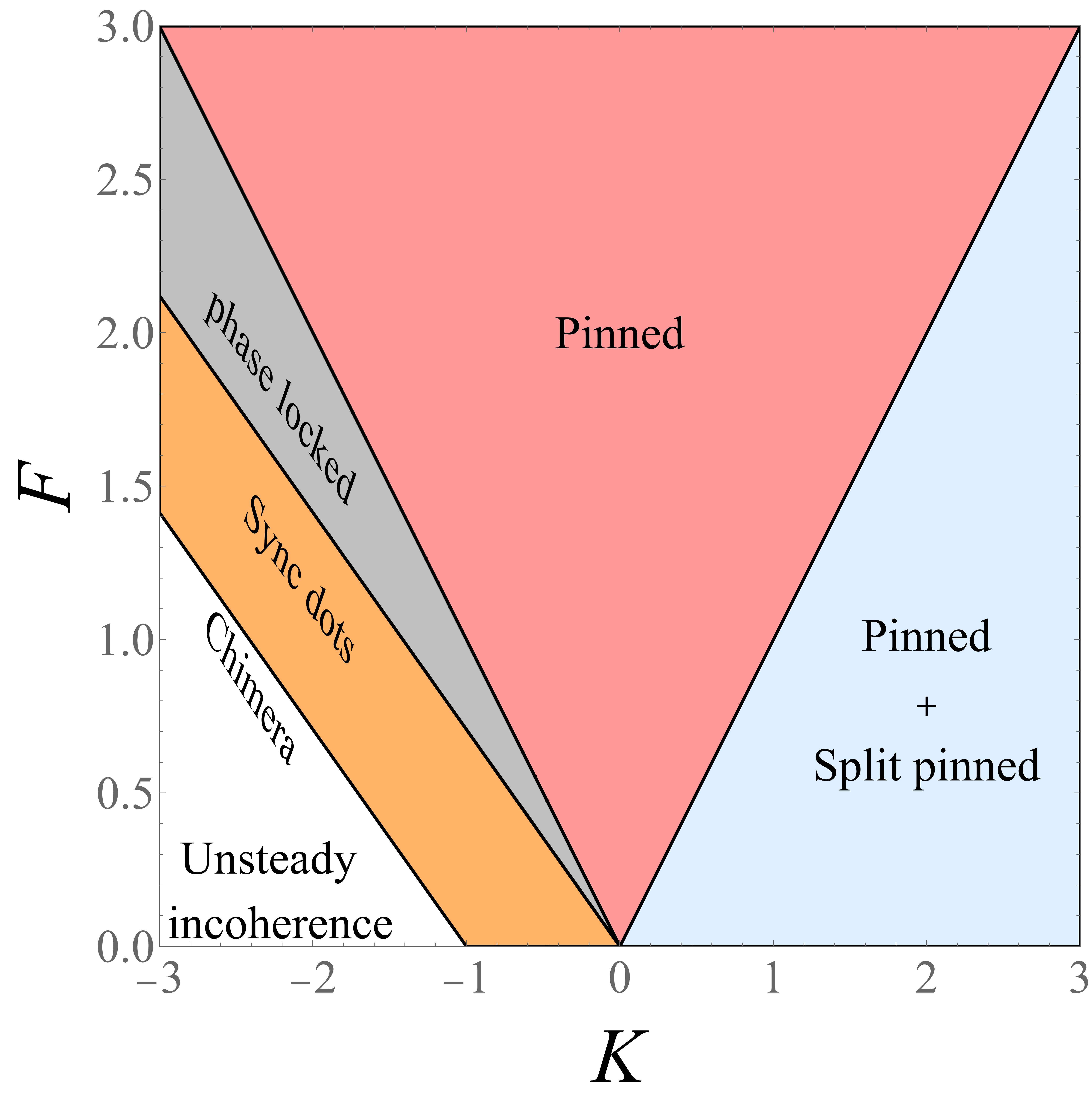}}
    \caption{Phase diagram in the $(K,F)$ plane. The solid black lines are the theoretically predicted boundaries for different states. }
    \label{bif}
\end{figure}

\par Figure~\ref{bif} shows where each state occurs in the $(K,F)$ plane, while Figs.~\ref{R_versus_F},~\ref{ops_vs_F} show the transition between the states as we vary the forcing amplitude $F$. The results are divided into two cases depending on the sign of the phase coupling $K$.

\par Figure~\ref{R_versus_F} shows what happens for positive phase coupling $K > 0$. The pinned state is stable for all $K>0$. The split pinned state, on the other hand, is stable only when $F < K$ (we will prove it analytically), when the pinning is weak relative to the phase coupling. Beyond a critical $F^*$, the pinning becomes the sole attractor. This makes sense when you consider that as $F/K \rightarrow \infty$, $\dot{\theta}_{i} \rightarrow -F \sin\theta_{i}$ which has just one fixed point.

\par Figure~\ref{ops_vs_F} shows the case of repulsive phase coupling $K<0$. For $K < K^*$, the system transition between five of the six collective states as $F$ is tuned from $0$. For small $F$, unsteady incoherence is realized. The large but negative phase coupling $|K| > F$ means the swarmalators want to maximize their phase differences, leading to a uniform distribution in $x,\theta$, which implies $S_{\pm} = 0$. Yet at the same time, the driving wants to pin the oscillators to $(x_i, \theta_i) = (C, 0)$, where $S_{\pm} = 1$. The result is an unsteady competition, which manifests as jumpy oscillations in the order parameters (Figures~\ref{ops_vs_t}(e),~\ref{ops_vs_t}(j)). 

\par As $F$ is turned up, the driving pushes the system faster and faster, and the mean speed $V$ increases monotonically (Figure~\ref{ops_vs_F}). This reaches a peak at some $F^*$, at which point the chimera is born. Here the $V$ declines monotonically, while $S_{\pm}, R, Q$ increase. The dynamics are quite subtle here. At first, the chimera is the lone attractor (grey shaded region), but for larger $F$, it coexists with the sync dots (red region). Figure \ref{Splus_vs_F} in Appendix-\ref{chim_dots_bistable} illustrates this bistability behavior. Then, for larger $F$, the sync dots become monostable, which eventually bifurcates into the phase locked state, which in turn bifurcates into the pinned state as $F \rightarrow \infty$. Finally, notice that for $K^* < K < 0$, only the sync dot $\rightarrow $ phase locked $\rightarrow$ pinned state transition sequence is realized (in Fig.\ref{bif}).

\par Next, we analyze the states, deriving the stability boundaries drawn as thick black lines in the phase diagram of Fig.~\ref{bif}.

%%%%%%%%%%%%%%%%%%%%%%%%%%%%%%%%%%%%
\section{Analysis}

\subsection{Pinned state}
In the pinned state, the fixed points are
\begin{align}
    x_i &= C, \; \theta_i= 0
\end{align}
for some constant $C$. The Jacobian $M_{pinned}$ evaluated at this point has a simple block structure
\begin{equation}
    M_{pinned} = \left[ 
\begin{array}{cc} 
  A_0 & 0 \\ 
  0 & A_1 \\
\end{array} 
\right], \label{Mpinned}
\end{equation}
where $A_{0}$ and $A_{1}$ are $N \times N$ blocks. Here,
\begin{equation}
A_0 = \begin{bmatrix}
    -\frac{N-1}{N}   & \frac{1}{N}  & \dots & \frac{1}{N}\\ \\
    \frac{1}{N}  & - \frac{N-1}{N} &  \dots & \frac{1}{N} \\ \\
    \hdotsfor{4} \\ \\ 
    \frac{1}{N}  & \frac{1}{N}  & \dots & - \frac{N-1}{N} 
\end{bmatrix}, \label{pinned_A0}
\end{equation}

\begin{equation}
A_1 = \begin{bmatrix}
    a_{0}   & a_{1}  & \dots & a_{N-1} \\
    a_{N-1}   &  a_{0}  &  \dots & a_{N-2}  \\
    \dots  & \dots & \dots \\
    a_{1}   & a_{2}   & \dots & a_{0}  
\end{bmatrix}, \label{pinned_A1}
\end{equation}
with
\begin{align}
    a_{0}=-K \frac{N-1}{N} - F , \\
    a_{1}=a_{2}=\dots=a_{N-1}= \frac{K}{N}. 
\end{align}
The eigenvalues of $M_{pinned}$ are the sum of the eigenvalues of $A_0$ and $A_1$. $A_0$ has one zero eigenvalue $\lambda_{A_0}$ = 0 stemming from the rotational symmetry of the model and $N-1$ stable eigennvalues $\lambda_{A_0} = -1$. The matrix $A_1$ is a circulant matrix and thus its eigenvalues are known exactly $\lambda_{A_{1}}\big\lvert_{j}=\sum\limits_{k=0}^{N-1} a_{k} \alpha^{kj}$ $(j=0,1,\dots, N-1)$ where $\alpha=e^{\frac{2\pi\mathcal{I}}{N}}$ is a primitive $n$-th root of unity and $\mathcal{I}$
is the imaginary unit.

\par Putting these altogether give
\begin{equation}
    \begin{array}{l}
     \lambda_0 = 0, \\ 
     \lambda_1 = -1, \\    
     \lambda_2 = -F,  \\
      \lambda_3 = -K -F,
    \end{array}
\end{equation}
with multiplicities $1$, $(N-1)$, $1$, $N-1$, respectively. This tells us that the pinned state dies via zero eigenvalue bifurcation at 
\begin{equation}\label{pinned_critical}
    F_c = - K,
\end{equation}
which is plotted in Fig.~\ref{bif}.
%%%%%%%%%%%%%%%%%%%%%%%%%%%%%%%%%%%%%%%%%%%%

\subsection{Split pinned state}
Here, the swarmalators form two clusters with fixed point $(x_1,\theta_1)=(C,0)$ and $(x_2,\theta_2)=(C+\pi,\pi)$. The Jacobian evaluated at the fixed points has a block structure similar to that of the pinned state, i.e.,
\begin{equation}
    M_{spilt \; pinned} = \left[ 
\begin{array}{cc} 
  A_0 & 0 \\ 
  0 & A_1 \\
\end{array} 
\right], \label{MSplitpinned}
\end{equation}
where $A_{0}$ is given as previously by Eq.~\eqref{pinned_A0}, and the elements of the block $A_{1}$ are given by
\begin{equation}
    \begin{array}{l}
         A_{1_{ij}}=\begin{cases}
             \frac{K}{N}, & i \neq j \\
             -F-\frac{N-1}{N}K & i=j, \; \text{and} \; 1\leq i \leq [\frac{N}{2}] \\
             F-\frac{N-1}{N}K & i=j, \; \text{and} \; [\frac{N}{2}]+1\leq i \leq N.
             
         \end{cases} 
    \end{array}
\end{equation}
The eigenvalues of $A_{1}$ are 
\begin{align}
    & \lambda_1 = -F-K, \\
    & \lambda_2 = F-K, \\
    & \lambda_{3,4} = \frac{1}{2} \left(-k \pm \sqrt{4 F^2+k^2}\right), 
\end{align}
with multiplicities $\frac{N}{2}-1$, $\frac{N}{2}-1$ and $1$, respectively. The eigenvalues of $A_{0}$ and $A_{1}$ altogether tells that the split pinned state exists for  
\begin{equation}\label{split_pinned_critical}
    F < K \; \text{for}\; K>0.
\end{equation}
The corresponding critical curve is plotted in Fig.~\ref{bif}.
%%%%%%%%%%%%%%%%%%%%%%%%%%%%%%%%%%%%%%%%%%%%
\subsection{Sync dots}
In the sync dots, swarmalators form two groups, and the number of swarmalators in each group depends on the initial conditions. The first group is defined by $(x_1, \theta_1) = (C, \theta^*)$, while the second is defined by $(C + \Delta x, -\theta^*)$. The constant $C$ is arbitrary and stems from the rotational symmetry in the $\dot{x}$ equations. The values $(\Delta x, \theta^*)$ can be found by substitution into the equation of motions \eqref{eoms-x} and \eqref{eoms-theta} as
\begin{align}
    \theta^* &= -\pi/4, \\
    \cos \Delta x &= -\frac{\sqrt{2} F}{K}. \\
\end{align}
The Jacobian of this state is given by 
\begin{equation}
    M_{sync \; dots} = \left[ 
\begin{array}{cc} 
  A & B \\ 
  C & D \\
\end{array} 
\right], \label{Mcluster}
\end{equation}
where $A,B, C, D$ are $N\times N$ blocks such that $C=KB$ and $D=-\frac{F}{\sqrt{2}}I+KA$. Here,
\begin{equation}
    A = \left[ 
\begin{array}{cc} 
  A_0 & 0 \\ 
  0 & A_0 \\
\end{array} 
\right],
\end{equation}

\begin{equation}
A_0 = \begin{bmatrix}
    -\frac{1}{N}(\frac{N}{2}-1)   & \frac{1}{N}  & \dots & \frac{1}{N}\\ \\
    \frac{1}{N}  & -\frac{1}{N}(\frac{N}{2}-1) &  \dots & \frac{1}{N} \\ \\
    \hdotsfor{4} \\ \\
    \frac{1}{N}  & \frac{1}{N}  & \dots & -\frac{1}{N}(\frac{N}{2}-1) 
\end{bmatrix}, 
\end{equation}

\begin{equation}
    B = \left[ 
\begin{array}{cc} 
  B_0 & B_1 \\ 
  B_1 & B_0 \\
\end{array} 
\right],
\end{equation}

\begin{equation}
B_0 = \begin{bmatrix}
    -\frac{1}{2}\sqrt{1-\frac{2F^2}{K^2}}   & 0  & \dots & 0\\
    0  & -\frac{1}{2}\sqrt{1-\frac{2F^2}{K^2}}&  \dots & 0 \\
    \hdotsfor{4} \\
    0  & 0  & \dots & -\frac{1}{2}\sqrt{1-\frac{2F^2}{K^2}} 
\end{bmatrix}, 
\end{equation}
and
\begin{equation}
B_1 = \begin{bmatrix}
    -\frac{1}{N}\sqrt{1-\frac{2F^2}{K^2}}   & -\frac{1}{N}\sqrt{1-\frac{2F^2}{K^2}}  & \dots & -\frac{1}{N}\sqrt{1-\frac{2F^2}{K^2}}\\
    -\frac{1}{N}\sqrt{1-\frac{2F^2}{K^2}}  & -\frac{1}{N}\sqrt{1-\frac{2F^2}{K^2}}&  \dots & -\frac{1}{N}\sqrt{1-\frac{2F^2}{K^2}} \\
    \hdotsfor{4} \\
    -\frac{1}{N}\sqrt{1-\frac{2F^2}{K^2}}  & -\frac{1}{N}\sqrt{1-\frac{2F^2}{K^2}}  & \dots & -\frac{1}{N}\sqrt{1-\frac{2F^2}{K^2}} 
\end{bmatrix}. 
\end{equation}
Now, the characteristic equation for the Jacobian $M_{sync \; dots}$ can be written by,
\begin{multline}
    det(M_{sync \; dots
}-\lambda I)= \\
    det \left[ (A-\lambda I) (D-\lambda I)-BC  \right]=det(G)=0
\end{multline}
Here, in the above equation, we used the fact that $A$ and $C$ commute with each other. Then, the matrix $G$ again can be represented as a block matrix as follows,
\begin{equation}
    G = \left[ 
\begin{array}{cc} 
  G_0 & G_1 \\ 
  G_1 & G_0 \\
\end{array} 
\right],
\end{equation}
where $G_1=cI$ with $c=\frac{K}{N}\left(1-\frac{2 F^2}{K^2}\right)$, and
\begin{equation}
G_0 = \begin{bmatrix}
    a   & b  & \dots & b\\
    b  & a&  \dots & b \\
    \hdotsfor{4} \\
    b  & b  & \dots & a 
\end{bmatrix}, 
\end{equation}
such that 
\begin{multline}
    a=\lambda ^2 +\frac{\left(\sqrt{2} F N+K (N-2)+N-2\right)}{2 N}\lambda \\ +\frac{\frac{F^2 (N+2)}{K}+\sqrt{2} F \left(\frac{N}{2}-1\right)-2 K}{2 N},
\end{multline} 
and
\begin{equation}
    b=-\frac{-2 F^2+\sqrt{2} F K+2 K (\lambda +\lambda  K+K)}{2 K N}.
\end{equation} 
We want the determinant of the matrix $G$, which is basically the product of its eigenvalues $\Tilde{\lambda}_{i}$, to be zero. Note that $\Tilde{\lambda}_{i}$ are different from the eigenvalue $\lambda_{i}$ of the Jacobian $M_{sync \; dots
}$. Now, the eigenvalues of $G$ are known exactly and given by 
$\Tilde{\lambda}_{1}=a-b$ with multiciplity $N-2$, $\Tilde{\lambda}_{2}=a+(\frac{N}{2}-1)b-\frac{N}{2}c$ with multiplicity $1$, and $\Tilde{\lambda}_{3}=a+(\frac{N}{2}-1)b+\frac{N}{2}c$ with multiciplity $1$. Plugging into the values of $a$, $b$ and $c$, we can obtain that the eigenvalues are independent of $N$ and are given by
\begin{equation}
    \begin{array}{l}
         \Tilde{\lambda}_{1}=\frac{2 F^2+\sqrt{2} F K (2 \lambda +1)+2 K \lambda  (K+2 \lambda +1)}{4 K}, \\
         \Tilde{\lambda}_{2}= \frac{2 F^2}{K}+\frac{F \lambda }{\sqrt{2}}+\lambda ^2-K, \\
         \Tilde{\lambda}_{3}= \frac{F \lambda }{\sqrt{2}}+\lambda ^2.
    \end{array}
\end{equation}
Since we have 
\begin{equation}
    det(G)= \prod_{i=1}^{N} \Tilde{\lambda}_{i}=0,
\end{equation}
each term $\Tilde{\lambda}_{i}$ must be zero, which eventually provides us the eigenvalues of the Jacobian $M_{sync \; dots}$ as follows,
\begin{equation}
    \begin{array}{l}
         \lambda_{1}=0, \\
        \lambda_{2}=-\frac{F}{\sqrt{2}}, \\
        \lambda_{3,4}=\frac{-K \left(\sqrt{2} F+K+1\right) \pm \sqrt{K \left(2 F^2 (K-4)+2 \sqrt{2} F (K-1) K+K (K+1)^2\right)}}{4 K}, \\
        \lambda_{5,6}= \frac{-F \sqrt{K}\pm\sqrt{ F^2 (K-16)+8 K^2}}{2 \sqrt{2K}}.
    \end{array} 
    \end{equation}
Now, we will use the eigenvalues of the Jacobian $M_{sync \; dots}$ to find the stability conditions. The eigenvalue $\lambda_{1}=0$ does not play any role in the stability of sync dots state. It is simply due to the rotational symmetry of the model. Similarly, the eigenvalue $\lambda_{2}$ is unimportant for the analysis as it has a negative real part for all the parameter regions of interest. However, the eigenvalues $\lambda_{3,4}$ and $\lambda_{5,6}$ play the role of stability indicators. Thus, the state becomes stable when
\begin{eqnarray}\label{sync_dots_critical}
     -\frac{(K+1)}{\sqrt{2}}<F<-\frac{K}{\sqrt{2}} \;\; \text{for} \;\;  K<0, 
\end{eqnarray}
which is plotted in Fig.~\ref{bif}.

%%%%%%%%%%%%%%%%%%%%%%%%%%%%%%%%%%%%%%%%%%%%%%%%%%%%%%%%%%%%%%%%%%%%%%%%%%%
\subsection{Phase locked}
The fixed points here take the form
\begin{eqnarray}
    x_{i}=C, \\
    \theta_{i} \in (-a,a),
\end{eqnarray}
where $C$ and $a$ are constants. Plugging these expressions into the equation of motions \eqref{eoms-x} and \eqref{eoms-theta} gives us
\begin{eqnarray}
    KR\sin{(\phi-\theta_{i})}-F\sin{\theta_{i}}=0,
\end{eqnarray}
where 
\begin{equation}
    Re^{\phi}=\frac{1}{N} \sum\limits_{j=1}^{N} e^{\mathrm{i}\theta_{j}}
\end{equation}
is the Kuramoto order parameter. Assuming $\phi=0$ without loss of generality, we have 
\begin{equation}
    KR=-F.
\end{equation}
Now, the state bifurcates from the sync dots state where the phases of the swarmalators are $\theta_{i}=-\frac{\pi}{4}$ for one group of the population and $\theta_{i}=\frac{\pi}{4}$ for another group. Putting the fixed point condition for the sync dots provides us with $R_{sync \; dots}=\frac{1}{\sqrt{2}}$. Thus, the phase locked state bifurcates from the sync dots through the critical curve 
\begin{eqnarray}
    F=-\frac{K}{\sqrt{2}},
\end{eqnarray}
which matches the stability boundary of the sync dots derived previously. Similarly, the pinned state is defined by $\theta_{i}=0$, which provides $R=1$. So, the pinned state bifurcates from the static coherent state through the critical curve $F=-K$, which matches the stability boundary of the pinned state.
\par Hence, the phase locked state exists when
\begin{eqnarray}\label{phase_locked_critical}
    -\frac{K}{\sqrt{2}}<F<-K \; \text{for}\; K<0.
\end{eqnarray}
The corresponding critical curve is delineated in Fig.~\ref{bif}.
%%%%%%%%%%%%%%%%%%%%%%%%%%%%%%%%%%%%%%%%%%%%%%%%%%%%%%%%%%%%%%%%%%
\subsection{Incoherence \& Chimeras}
Being unsteady, we were unable to analyze these states. We hope future researchers can shed some analytic light here, especially on the swarmalator chimera.

%%%%%%%%%%%%%%%%%%%%%%%%%%%%%%%%%%%%%%%%%%%%%%%%%%%%%%%%%%%%%%%%%%
\section{Discussion}
We studied a simple model of forced swarmalators which exhibited a variety of collective states. We analyze all static states and provide phase boundaries for all but one of the states. A surprise finding was the swarmalator chimera, which has never been seen before.

\par A natural puzzle for future work is to analyze this chimera. Why do the incoherent swarmalators run in a peanut-shaped loop? What controls the shape? What is the critical driving $F_c$ at which the chimera grows out of unsteady incoherence? We know it is born via a Hopf bifurcation from the sync dots, but what is the bifurcation type on the incoherent side? We suspect it might be a global bifurcation since it separates two unsteady states, but further investigations are needed to back up this intuition. Chimeras usually arise when there is a phase frustration term $\alpha$ in the inter-element coupling $\sin(\theta_j - \theta_i - \alpha)$ \cite{kuramoto2002coexistence,abrams2004chimera}. Yet that doesn't appear to happen here. Why? Is this exclusive to swarmalators, or might there be varieties of regular "oscillator chimeras" that sustain themselves without phase frustration $\alpha=0$? We hope other researchers will join us in grappling with these questions in future work.

\par Another avenue to explore would be to relax some of the idealizations of the 1D swarmalator model we studied. For simplicity's sake (we were aiming, recall, for maximum tractability), we analyzed the simplest model of forced swarmalators we could think of: one where the spatial motion was confined to one periodic dimension, the swarmalators were identical, the forcing was resonant $(\Omega = \omega)$ and fixed position coupling. Effects like non-resonant driving, environmental noise, heterogeneous natural frequencies, local coupling, random coupling, delayed coupling, motion in two spatial dimensions, and so on could be added back in to make the model more faithful to real forced swarmalators systems. 

Many of these and other effects have been studied in swarmalator models in the absence of forcing \cite{o2022collective,yoon2022sync,hong2023swarmalators,lee2021collective,o2018ring,o2022swarmalators,hao2023attractive,o2023solvable,blum2024swarmalators,lizarraga2023synchronization,o2018ring,ceron2023diverse,ghosh2023antiphase,schilcher2021swarmalators,kongni2023phase,sar2023pinning,sar2023swarmalators,degond2023topological,ansarinasab2024spatial,smith2024swarmalators,yadav2024exotic,sar2024solvable,schilcher2023radii,adorjani2023motility,anwar2024collective}. These analyses may be useful as starting points. In particular, the solvable 2D model of swarmalators recently presented in \cite{o2023solvable} would be a nice play to start studying the more realistic case of 2D or even 3D colloids. The model has the same form as the 1D model presented here and so inherits its tractability (i.e. we expect it may analyzed with the same methods we used here).\\\\

%\vspace{4cm}

\appendix
\section{Compact support of the phases of swarmalators in phase locked state} \label{phase_locked_histogram}
\begin{figure}[hpt]
	\centering
	\includegraphics[width = 0.7 \columnwidth]{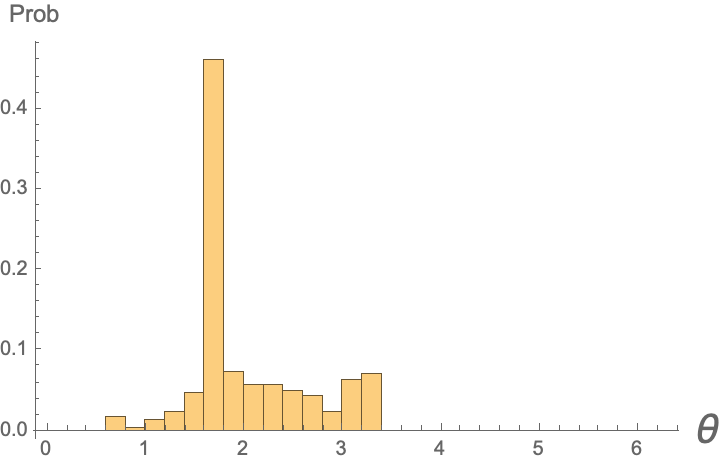}
	\caption{Histogram of phases in the phase locked state. The phases have compact support $\theta \in (-a, a)$ for some $a<\pi$. The width of the support $a$ depends on the system parameters.}
	\label{phase_locked}
\end{figure}

Figure~\ref{phase_locked} shows the phase distribution in the phase locked state has compact support.

\section{Coexistence of sync dots and chimera state} \label{chim_dots_bistable}
\begin{figure}[htp]
    \centering
    {\includegraphics[width=\columnwidth]{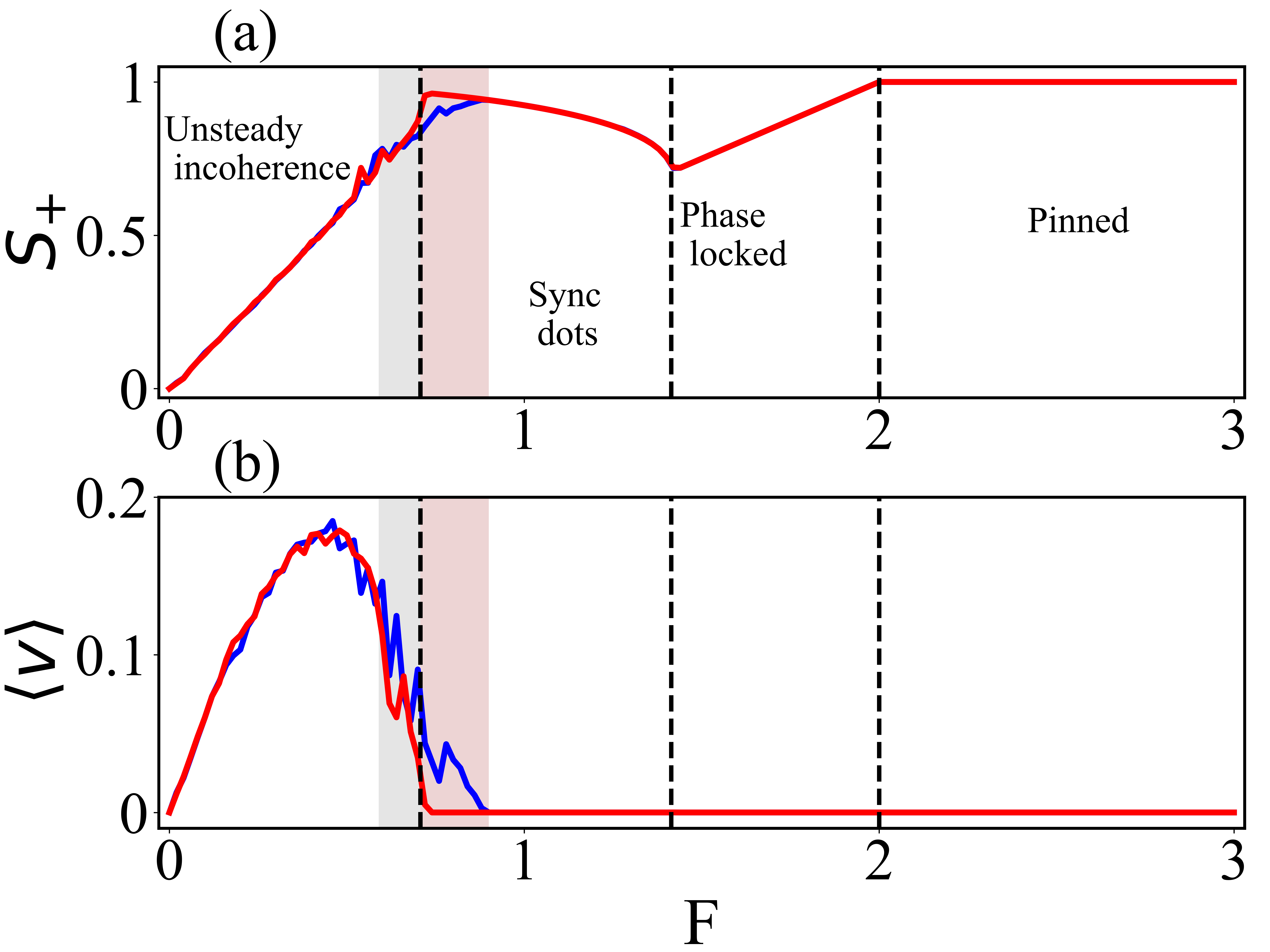}}
    \caption{ (a) $S_{+}(F)$, and (b) $\langle V \rangle(F)$ for $K=-2$. Other simulation parameters used are $(N, T)=(1000,1500)$. The dotted black vertical lines are the theoretical boundaries of the static states. The blue and red curves in both panels indicate the results for two different initial conditions: the blue curves are for the choice of initial conditions uniformly at random from the square of length $2\pi$, while the red curve corresponds to the result with initial conditions drawn uniformly at random from the interval $(0,0.1\pi)$. It is observable that within the shaded red region, both sync dots and chimera states coexist. Here, the other order parameters are not plotted since they do not provide any additional information.}
    \label{Splus_vs_F}
\end{figure}

For the case of repulsive phase coupling $K<0$, as $F$ is turned up, one can observe the coexistence of chimera and the sync dots within a small parameter region (red shaded region in Fig.~\ref{ops_vs_F}). Here, both the states emerge depending on the choice of initial positions and phases of the swarmalators. Specifically, the chimera state occurs when the initial conditions are drawn uniformly at random from the square of length $2\pi$. In contrast, the sync dots emerge when the initial conditions are drawn from relatively smaller squares, i.e., when the initial positions and phases of the swarmalators are very close to one another. Figure~\ref{Splus_vs_F} illustrates this bistable nature by plotting the rainbow order parameter $S_{+}(F)$ and the mean velocity $V$ for the aforementioned two different sets of initial conditions. The blue curves represent the result for initial conditions drawn from the square of length $2\pi$, revealing an increasing trend of $S_{+}(F)$ and a non-zero decreasing trend of mean velocity $\langle V \rangle(F)$ within the red-shaded region. This suggests the occurrence of the chimera state. Whereas the red curves depict the results for closer initials, demonstrating a decreasing pattern of $S_{+}(F)$ and zero mean velocity, suggesting the emergence of the sync dots. 

\bibliographystyle{apsrev}
\bibliography{ref.bib}

\begin{thebibliography}{52}
\expandafter\ifx\csname natexlab\endcsname\relax\def\natexlab#1{#1}\fi
\expandafter\ifx\csname bibnamefont\endcsname\relax
  \def\bibnamefont#1{#1}\fi
\expandafter\ifx\csname bibfnamefont\endcsname\relax
  \def\bibfnamefont#1{#1}\fi
\expandafter\ifx\csname citenamefont\endcsname\relax
  \def\citenamefont#1{#1}\fi
\expandafter\ifx\csname url\endcsname\relax
  \def\url#1{\texttt{#1}}\fi
\expandafter\ifx\csname urlprefix\endcsname\relax\def\urlprefix{URL }\fi
\providecommand{\bibinfo}[2]{#2}
\providecommand{\eprint}[2][]{\url{#2}}

\bibitem[{\citenamefont{Buck}(1988)}]{buck1988synchronous}
\bibinfo{author}{\bibfnamefont{J.}~\bibnamefont{Buck}}, \bibinfo{journal}{The
  Quarterly Review of Biology} \textbf{\bibinfo{volume}{63}},
  \bibinfo{pages}{265} (\bibinfo{year}{1988}).

\bibitem[{\citenamefont{Kozyreff et~al.}(2000)\citenamefont{Kozyreff,
  Vladimirov, and Mandel}}]{kozyreff2000global}
\bibinfo{author}{\bibfnamefont{G.}~\bibnamefont{Kozyreff}},
  \bibinfo{author}{\bibfnamefont{A.}~\bibnamefont{Vladimirov}},
  \bibnamefont{and} \bibinfo{author}{\bibfnamefont{P.}~\bibnamefont{Mandel}},
  \bibinfo{journal}{Physical Review Letters} \textbf{\bibinfo{volume}{85}},
  \bibinfo{pages}{3809} (\bibinfo{year}{2000}).

\bibitem[{\citenamefont{Peskin}(1975)}]{peskin75}
\bibinfo{author}{\bibfnamefont{C.~S.} \bibnamefont{Peskin}},
  \emph{\bibinfo{title}{Mathematical Aspects of Heart Physiology}}
  (\bibinfo{publisher}{Courant Institute of Mathematical Sciences},
  \bibinfo{address}{New York}, \bibinfo{year}{1975}), pp.
  \bibinfo{pages}{268--278}.

\bibitem[{\citenamefont{Bialek et~al.}(2012)\citenamefont{Bialek, Cavagna,
  Giardina, Mora, Silvestri, Viale, and Walczak}}]{bialek2012statistical}
\bibinfo{author}{\bibfnamefont{W.}~\bibnamefont{Bialek}},
  \bibinfo{author}{\bibfnamefont{A.}~\bibnamefont{Cavagna}},
  \bibinfo{author}{\bibfnamefont{I.}~\bibnamefont{Giardina}},
  \bibinfo{author}{\bibfnamefont{T.}~\bibnamefont{Mora}},
  \bibinfo{author}{\bibfnamefont{E.}~\bibnamefont{Silvestri}},
  \bibinfo{author}{\bibfnamefont{M.}~\bibnamefont{Viale}}, \bibnamefont{and}
  \bibinfo{author}{\bibfnamefont{A.~M.} \bibnamefont{Walczak}},
  \bibinfo{journal}{Proceedings of the National Academy of Sciences}
  \textbf{\bibinfo{volume}{109}}, \bibinfo{pages}{4786} (\bibinfo{year}{2012}).

\bibitem[{\citenamefont{Katz et~al.}(2011)\citenamefont{Katz, Tunstr{\o}m,
  Ioannou, Huepe, and Couzin}}]{katz2011inferring}
\bibinfo{author}{\bibfnamefont{Y.}~\bibnamefont{Katz}},
  \bibinfo{author}{\bibfnamefont{K.}~\bibnamefont{Tunstr{\o}m}},
  \bibinfo{author}{\bibfnamefont{C.~C.} \bibnamefont{Ioannou}},
  \bibinfo{author}{\bibfnamefont{C.}~\bibnamefont{Huepe}}, \bibnamefont{and}
  \bibinfo{author}{\bibfnamefont{I.~D.} \bibnamefont{Couzin}},
  \bibinfo{journal}{Proceedings of the National Academy of Sciences}
  \textbf{\bibinfo{volume}{108}}, \bibinfo{pages}{18720}
  (\bibinfo{year}{2011}).

\bibitem[{\citenamefont{Garcimart\'{\i}n
  et~al.}(2015)\citenamefont{Garcimart\'{\i}n, Pastor, Ferrer, Ramos,
  Mart\'{\i}n-G\'omez, and Zuriguel}}]{garcimartin2015}
\bibinfo{author}{\bibfnamefont{A.}~\bibnamefont{Garcimart\'{\i}n}},
  \bibinfo{author}{\bibfnamefont{J.~M.} \bibnamefont{Pastor}},
  \bibinfo{author}{\bibfnamefont{L.~M.} \bibnamefont{Ferrer}},
  \bibinfo{author}{\bibfnamefont{J.~J.} \bibnamefont{Ramos}},
  \bibinfo{author}{\bibfnamefont{C.}~\bibnamefont{Mart\'{\i}n-G\'omez}},
  \bibnamefont{and} \bibinfo{author}{\bibfnamefont{I.}~\bibnamefont{Zuriguel}},
  \bibinfo{journal}{Physical Review E} \textbf{\bibinfo{volume}{91}},
  \bibinfo{pages}{022808} (\bibinfo{year}{2015}).

\bibitem[{\citenamefont{Tamm et~al.}(1975)\citenamefont{Tamm, Sonneborn, and
  Dippell}}]{tamm1975role}
\bibinfo{author}{\bibfnamefont{S.~L.} \bibnamefont{Tamm}},
  \bibinfo{author}{\bibfnamefont{T.}~\bibnamefont{Sonneborn}},
  \bibnamefont{and} \bibinfo{author}{\bibfnamefont{R.~V.}
  \bibnamefont{Dippell}}, \bibinfo{journal}{The Journal of Cell Biology}
  \textbf{\bibinfo{volume}{64}}, \bibinfo{pages}{98} (\bibinfo{year}{1975}).

\bibitem[{\citenamefont{Verberck}(2022)}]{verberck2022wavy}
\bibinfo{author}{\bibfnamefont{B.}~\bibnamefont{Verberck}},
  \bibinfo{journal}{Nature Physics} \textbf{\bibinfo{volume}{18}},
  \bibinfo{pages}{131} (\bibinfo{year}{2022}).

\bibitem[{\citenamefont{Belovs et~al.}(2017)\citenamefont{Belovs,
  Livanovi{\v{c}}s, and C{\=e}bers}}]{belovs2017synchronized}
\bibinfo{author}{\bibfnamefont{M.}~\bibnamefont{Belovs}},
  \bibinfo{author}{\bibfnamefont{R.}~\bibnamefont{Livanovi{\v{c}}s}},
  \bibnamefont{and}
  \bibinfo{author}{\bibfnamefont{A.}~\bibnamefont{C{\=e}bers}},
  \bibinfo{journal}{Physical Review E} \textbf{\bibinfo{volume}{96}},
  \bibinfo{pages}{042408} (\bibinfo{year}{2017}).

\bibitem[{\citenamefont{Yan et~al.}(2012)\citenamefont{Yan, Bloom, Bae,
  Luijten, and Granick}}]{yan2012linking}
\bibinfo{author}{\bibfnamefont{J.}~\bibnamefont{Yan}},
  \bibinfo{author}{\bibfnamefont{M.}~\bibnamefont{Bloom}},
  \bibinfo{author}{\bibfnamefont{S.~C.} \bibnamefont{Bae}},
  \bibinfo{author}{\bibfnamefont{E.}~\bibnamefont{Luijten}}, \bibnamefont{and}
  \bibinfo{author}{\bibfnamefont{S.}~\bibnamefont{Granick}},
  \bibinfo{journal}{Nature} \textbf{\bibinfo{volume}{491}},
  \bibinfo{pages}{578} (\bibinfo{year}{2012}).

\bibitem[{\citenamefont{Hwang et~al.}(2020)\citenamefont{Hwang, Nguyen,
  Bhaskar, Yoon, Klaiber, Lee, Glotzer, and Lahann}}]{hwang2020cooperative}
\bibinfo{author}{\bibfnamefont{S.}~\bibnamefont{Hwang}},
  \bibinfo{author}{\bibfnamefont{T.~D.} \bibnamefont{Nguyen}},
  \bibinfo{author}{\bibfnamefont{S.}~\bibnamefont{Bhaskar}},
  \bibinfo{author}{\bibfnamefont{J.}~\bibnamefont{Yoon}},
  \bibinfo{author}{\bibfnamefont{M.}~\bibnamefont{Klaiber}},
  \bibinfo{author}{\bibfnamefont{K.~J.} \bibnamefont{Lee}},
  \bibinfo{author}{\bibfnamefont{S.~C.} \bibnamefont{Glotzer}},
  \bibnamefont{and} \bibinfo{author}{\bibfnamefont{J.}~\bibnamefont{Lahann}},
  \bibinfo{journal}{Advanced Functional Materials}
  \textbf{\bibinfo{volume}{30}}, \bibinfo{pages}{1907865}
  (\bibinfo{year}{2020}).

\bibitem[{\citenamefont{Zhang et~al.}(2020)\citenamefont{Zhang, Sokolov, and
  Snezhko}}]{zhang2020reconfigurable}
\bibinfo{author}{\bibfnamefont{B.}~\bibnamefont{Zhang}},
  \bibinfo{author}{\bibfnamefont{A.}~\bibnamefont{Sokolov}}, \bibnamefont{and}
  \bibinfo{author}{\bibfnamefont{A.}~\bibnamefont{Snezhko}},
  \bibinfo{journal}{Nature Communications} \textbf{\bibinfo{volume}{11}},
  \bibinfo{pages}{1} (\bibinfo{year}{2020}).

\bibitem[{\citenamefont{Bricard et~al.}(2015)\citenamefont{Bricard, Caussin,
  Das, Savoie, Chikkadi, Shitara, Chepizhko, Peruani, Saintillan, and
  Bartolo}}]{bricard2015emergent}
\bibinfo{author}{\bibfnamefont{A.}~\bibnamefont{Bricard}},
  \bibinfo{author}{\bibfnamefont{J.-B.} \bibnamefont{Caussin}},
  \bibinfo{author}{\bibfnamefont{D.}~\bibnamefont{Das}},
  \bibinfo{author}{\bibfnamefont{C.}~\bibnamefont{Savoie}},
  \bibinfo{author}{\bibfnamefont{V.}~\bibnamefont{Chikkadi}},
  \bibinfo{author}{\bibfnamefont{K.}~\bibnamefont{Shitara}},
  \bibinfo{author}{\bibfnamefont{O.}~\bibnamefont{Chepizhko}},
  \bibinfo{author}{\bibfnamefont{F.}~\bibnamefont{Peruani}},
  \bibinfo{author}{\bibfnamefont{D.}~\bibnamefont{Saintillan}},
  \bibnamefont{and} \bibinfo{author}{\bibfnamefont{D.}~\bibnamefont{Bartolo}},
  \bibinfo{journal}{Nature Communications} \textbf{\bibinfo{volume}{6}},
  \bibinfo{pages}{1} (\bibinfo{year}{2015}).

\bibitem[{\citenamefont{Manna et~al.}(2021)\citenamefont{Manna, Shklyaev, and
  Balazs}}]{manna2021chemical}
\bibinfo{author}{\bibfnamefont{R.~K.} \bibnamefont{Manna}},
  \bibinfo{author}{\bibfnamefont{O.~E.} \bibnamefont{Shklyaev}},
  \bibnamefont{and} \bibinfo{author}{\bibfnamefont{A.~C.}
  \bibnamefont{Balazs}}, \bibinfo{journal}{Proceedings of the National Academy
  of Sciences} \textbf{\bibinfo{volume}{118}} (\bibinfo{year}{2021}).

\bibitem[{\citenamefont{Riedel et~al.}(2005)\citenamefont{Riedel, Kruse, and
  Howard}}]{riedel2005self}
\bibinfo{author}{\bibfnamefont{I.~H.} \bibnamefont{Riedel}},
  \bibinfo{author}{\bibfnamefont{K.}~\bibnamefont{Kruse}}, \bibnamefont{and}
  \bibinfo{author}{\bibfnamefont{J.}~\bibnamefont{Howard}},
  \bibinfo{journal}{Science} \textbf{\bibinfo{volume}{309}},
  \bibinfo{pages}{300} (\bibinfo{year}{2005}).

\bibitem[{\citenamefont{Yang et~al.}(2008)\citenamefont{Yang, Elgeti, and
  Gompper}}]{yang2008cooperation}
\bibinfo{author}{\bibfnamefont{Y.}~\bibnamefont{Yang}},
  \bibinfo{author}{\bibfnamefont{J.}~\bibnamefont{Elgeti}}, \bibnamefont{and}
  \bibinfo{author}{\bibfnamefont{G.}~\bibnamefont{Gompper}},
  \bibinfo{journal}{Physical Review E} \textbf{\bibinfo{volume}{78}},
  \bibinfo{pages}{061903} (\bibinfo{year}{2008}).

\bibitem[{\citenamefont{O'Keeffe et~al.}(2017)\citenamefont{O'Keeffe, Hong, and
  Strogatz}}]{o2017oscillators}
\bibinfo{author}{\bibfnamefont{K.~P.} \bibnamefont{O'Keeffe}},
  \bibinfo{author}{\bibfnamefont{H.}~\bibnamefont{Hong}}, \bibnamefont{and}
  \bibinfo{author}{\bibfnamefont{S.~H.} \bibnamefont{Strogatz}},
  \bibinfo{journal}{Nature Communications} \textbf{\bibinfo{volume}{8}},
  \bibinfo{pages}{1} (\bibinfo{year}{2017}).

\bibitem[{\citenamefont{Sar et~al.}(2024)\citenamefont{Sar, Ghosh, and
  O'Keeffe}}]{sar2024solvable}
\bibinfo{author}{\bibfnamefont{G.~K.} \bibnamefont{Sar}},
  \bibinfo{author}{\bibfnamefont{D.}~\bibnamefont{Ghosh}}, \bibnamefont{and}
  \bibinfo{author}{\bibfnamefont{K.}~\bibnamefont{O'Keeffe}},
  \bibinfo{journal}{Physical Review E} \textbf{\bibinfo{volume}{109}},
  \bibinfo{pages}{044603} (\bibinfo{year}{2024}).

\bibitem[{\citenamefont{Sar et~al.}(2023{\natexlab{a}})\citenamefont{Sar,
  Ghosh, and O'Keeffe}}]{sar2023pinning}
\bibinfo{author}{\bibfnamefont{G.~K.} \bibnamefont{Sar}},
  \bibinfo{author}{\bibfnamefont{D.}~\bibnamefont{Ghosh}}, \bibnamefont{and}
  \bibinfo{author}{\bibfnamefont{K.}~\bibnamefont{O'Keeffe}},
  \bibinfo{journal}{Physical Review E} \textbf{\bibinfo{volume}{107}},
  \bibinfo{pages}{024215} (\bibinfo{year}{2023}{\natexlab{a}}).

\bibitem[{\citenamefont{Blum et~al.}(2024)\citenamefont{Blum, Li, O'Keeffe, and
  Kogan}}]{blum2024swarmalators}
\bibinfo{author}{\bibfnamefont{N.}~\bibnamefont{Blum}},
  \bibinfo{author}{\bibfnamefont{A.}~\bibnamefont{Li}},
  \bibinfo{author}{\bibfnamefont{K.}~\bibnamefont{O'Keeffe}}, \bibnamefont{and}
  \bibinfo{author}{\bibfnamefont{O.}~\bibnamefont{Kogan}},
  \bibinfo{journal}{Physical Review E} \textbf{\bibinfo{volume}{109}},
  \bibinfo{pages}{014205} (\bibinfo{year}{2024}).

\bibitem[{\citenamefont{Lee et~al.}(2021)\citenamefont{Lee, Yeo, and
  Hong}}]{lee2021collective}
\bibinfo{author}{\bibfnamefont{H.~K.} \bibnamefont{Lee}},
  \bibinfo{author}{\bibfnamefont{K.}~\bibnamefont{Yeo}}, \bibnamefont{and}
  \bibinfo{author}{\bibfnamefont{H.}~\bibnamefont{Hong}},
  \bibinfo{journal}{Chaos: An Interdisciplinary Journal of Nonlinear Science}
  \textbf{\bibinfo{volume}{31}}, \bibinfo{pages}{033134}
  (\bibinfo{year}{2021}).

\bibitem[{\citenamefont{Hong}(2018)}]{hong2018active}
\bibinfo{author}{\bibfnamefont{H.}~\bibnamefont{Hong}},
  \bibinfo{journal}{Chaos: An Interdisciplinary Journal of Nonlinear Science}
  \textbf{\bibinfo{volume}{28}}, \bibinfo{pages}{103112}
  (\bibinfo{year}{2018}).

\bibitem[{\citenamefont{McLennan-Smith
  et~al.}(2020)\citenamefont{McLennan-Smith, Roberts, and
  Sidhu}}]{mclennan2020emergent}
\bibinfo{author}{\bibfnamefont{T.~A.} \bibnamefont{McLennan-Smith}},
  \bibinfo{author}{\bibfnamefont{D.~O.} \bibnamefont{Roberts}},
  \bibnamefont{and} \bibinfo{author}{\bibfnamefont{H.~S.} \bibnamefont{Sidhu}},
  \bibinfo{journal}{Physical Review E} \textbf{\bibinfo{volume}{102}},
  \bibinfo{pages}{032607} (\bibinfo{year}{2020}).

\bibitem[{\citenamefont{Jim{\'e}nez-Morales}(2020)}]{jimenez2020oscillatory}
\bibinfo{author}{\bibfnamefont{F.}~\bibnamefont{Jim{\'e}nez-Morales}},
  \bibinfo{journal}{Physical Review E} \textbf{\bibinfo{volume}{101}},
  \bibinfo{pages}{062202} (\bibinfo{year}{2020}).

\bibitem[{\citenamefont{Liz{\'a}rraga and
  de~Aguiar}(2023)}]{lizarraga2023synchronization}
\bibinfo{author}{\bibfnamefont{J.~U.} \bibnamefont{Liz{\'a}rraga}}
  \bibnamefont{and} \bibinfo{author}{\bibfnamefont{M.~A.}
  \bibnamefont{de~Aguiar}}, \bibinfo{journal}{Physical Review E}
  \textbf{\bibinfo{volume}{108}}, \bibinfo{pages}{024212}
  (\bibinfo{year}{2023}).

\bibitem[{\citenamefont{Ghosh et~al.}(2023)\citenamefont{Ghosh, Sar, Majhi, and
  Ghosh}}]{ghosh2023antiphase}
\bibinfo{author}{\bibfnamefont{S.}~\bibnamefont{Ghosh}},
  \bibinfo{author}{\bibfnamefont{G.~K.} \bibnamefont{Sar}},
  \bibinfo{author}{\bibfnamefont{S.}~\bibnamefont{Majhi}}, \bibnamefont{and}
  \bibinfo{author}{\bibfnamefont{D.}~\bibnamefont{Ghosh}},
  \bibinfo{journal}{Physical Review E} \textbf{\bibinfo{volume}{108}},
  \bibinfo{pages}{034217} (\bibinfo{year}{2023}).

\bibitem[{\citenamefont{Yadav et~al.}(2024)\citenamefont{Yadav, Chandrasekar,
  Zou, Kurths, and Senthilkumar}}]{yadav2024exotic}
\bibinfo{author}{\bibfnamefont{A.}~\bibnamefont{Yadav}},
  \bibinfo{author}{\bibfnamefont{V.}~\bibnamefont{Chandrasekar}},
  \bibinfo{author}{\bibfnamefont{W.}~\bibnamefont{Zou}},
  \bibinfo{author}{\bibfnamefont{J.}~\bibnamefont{Kurths}}, \bibnamefont{and}
  \bibinfo{author}{\bibfnamefont{D.}~\bibnamefont{Senthilkumar}},
  \bibinfo{journal}{Physical Review E} \textbf{\bibinfo{volume}{109}},
  \bibinfo{pages}{044212} (\bibinfo{year}{2024}).

\bibitem[{\citenamefont{Kongni et~al.}(2023)\citenamefont{Kongni, Nguefoue,
  Njougouo, Louodop, Ferreira, Tchitnga, and Cerdeira}}]{kongni2023phase}
\bibinfo{author}{\bibfnamefont{S.~J.} \bibnamefont{Kongni}},
  \bibinfo{author}{\bibfnamefont{V.}~\bibnamefont{Nguefoue}},
  \bibinfo{author}{\bibfnamefont{T.}~\bibnamefont{Njougouo}},
  \bibinfo{author}{\bibfnamefont{P.}~\bibnamefont{Louodop}},
  \bibinfo{author}{\bibfnamefont{F.~F.} \bibnamefont{Ferreira}},
  \bibinfo{author}{\bibfnamefont{R.}~\bibnamefont{Tchitnga}}, \bibnamefont{and}
  \bibinfo{author}{\bibfnamefont{H.~A.} \bibnamefont{Cerdeira}},
  \bibinfo{journal}{Physical Review E} \textbf{\bibinfo{volume}{108}},
  \bibinfo{pages}{034303} (\bibinfo{year}{2023}).

\bibitem[{\citenamefont{Anwar et~al.}(2024)\citenamefont{Anwar, Sar, Perc, and
  Ghosh}}]{anwar2024collective}
\bibinfo{author}{\bibfnamefont{M.~S.} \bibnamefont{Anwar}},
  \bibinfo{author}{\bibfnamefont{G.~K.} \bibnamefont{Sar}},
  \bibinfo{author}{\bibfnamefont{M.}~\bibnamefont{Perc}}, \bibnamefont{and}
  \bibinfo{author}{\bibfnamefont{D.}~\bibnamefont{Ghosh}},
  \bibinfo{journal}{Communications Physics} \textbf{\bibinfo{volume}{7}},
  \bibinfo{pages}{59} (\bibinfo{year}{2024}).

\bibitem[{\citenamefont{O'Keeffe et~al.}(2018)\citenamefont{O'Keeffe, Evers,
  and Kolokolnikov}}]{o2018ring}
\bibinfo{author}{\bibfnamefont{K.~P.} \bibnamefont{O'Keeffe}},
  \bibinfo{author}{\bibfnamefont{J.~H.} \bibnamefont{Evers}}, \bibnamefont{and}
  \bibinfo{author}{\bibfnamefont{T.}~\bibnamefont{Kolokolnikov}},
  \bibinfo{journal}{Physical Review E} \textbf{\bibinfo{volume}{98}},
  \bibinfo{pages}{022203} (\bibinfo{year}{2018}).

\bibitem[{\citenamefont{Smith}(2024)}]{smith2024swarmalators}
\bibinfo{author}{\bibfnamefont{L.~D.} \bibnamefont{Smith}},
  \bibinfo{journal}{SIAM Journal on Applied Dynamical Systems}
  \textbf{\bibinfo{volume}{23}}, \bibinfo{pages}{1133} (\bibinfo{year}{2024}).

\bibitem[{\citenamefont{Ansarinasab et~al.}(2024)\citenamefont{Ansarinasab,
  Nazarimehr, Ghassemi, Ghosh, and Jafari}}]{ansarinasab2024spatial}
\bibinfo{author}{\bibfnamefont{S.}~\bibnamefont{Ansarinasab}},
  \bibinfo{author}{\bibfnamefont{F.}~\bibnamefont{Nazarimehr}},
  \bibinfo{author}{\bibfnamefont{F.}~\bibnamefont{Ghassemi}},
  \bibinfo{author}{\bibfnamefont{D.}~\bibnamefont{Ghosh}}, \bibnamefont{and}
  \bibinfo{author}{\bibfnamefont{S.}~\bibnamefont{Jafari}},
  \bibinfo{journal}{Applied Mathematics and Computation}
  \textbf{\bibinfo{volume}{468}}, \bibinfo{pages}{128508}
  (\bibinfo{year}{2024}).

\bibitem[{\citenamefont{Schilcher et~al.}(2021)\citenamefont{Schilcher,
  Schmidt, Vogell, and Bettstetter}}]{schilcher2021swarmalators}
\bibinfo{author}{\bibfnamefont{U.}~\bibnamefont{Schilcher}},
  \bibinfo{author}{\bibfnamefont{J.~F.} \bibnamefont{Schmidt}},
  \bibinfo{author}{\bibfnamefont{A.}~\bibnamefont{Vogell}}, \bibnamefont{and}
  \bibinfo{author}{\bibfnamefont{C.}~\bibnamefont{Bettstetter}}, in
  \emph{\bibinfo{booktitle}{2021 IEEE International Conference on Autonomic
  Computing and Self-Organizing Systems (ACSOS)}}
  (\bibinfo{organization}{IEEE}, \bibinfo{year}{2021}), pp.
  \bibinfo{pages}{90--99}.

\bibitem[{\citenamefont{Degond et~al.}(2022)\citenamefont{Degond, Diez, and
  Walczak}}]{degond2022topological}
\bibinfo{author}{\bibfnamefont{P.}~\bibnamefont{Degond}},
  \bibinfo{author}{\bibfnamefont{A.}~\bibnamefont{Diez}}, \bibnamefont{and}
  \bibinfo{author}{\bibfnamefont{A.}~\bibnamefont{Walczak}},
  \bibinfo{journal}{Analysis and Applications} \textbf{\bibinfo{volume}{20}},
  \bibinfo{pages}{1215} (\bibinfo{year}{2022}).

\bibitem[{\citenamefont{Liz{\'a}rraga et~al.}(2024)\citenamefont{Liz{\'a}rraga,
  O'Keeffe, and de~Aguiar}}]{lizarraga2024order}
\bibinfo{author}{\bibfnamefont{J.~U.} \bibnamefont{Liz{\'a}rraga}},
  \bibinfo{author}{\bibfnamefont{K.~P.} \bibnamefont{O'Keeffe}},
  \bibnamefont{and} \bibinfo{author}{\bibfnamefont{M.~A.}
  \bibnamefont{de~Aguiar}}, \bibinfo{journal}{Physical Review E}
  \textbf{\bibinfo{volume}{109}}, \bibinfo{pages}{044209}
  (\bibinfo{year}{2024}).

\bibitem[{\citenamefont{Gong et~al.}(2024)\citenamefont{Gong, Zhou, and
  Huang}}]{gong2024approximating}
\bibinfo{author}{\bibfnamefont{Z.}~\bibnamefont{Gong}},
  \bibinfo{author}{\bibfnamefont{J.}~\bibnamefont{Zhou}}, \bibnamefont{and}
  \bibinfo{author}{\bibfnamefont{M.}~\bibnamefont{Huang}},
  \bibinfo{journal}{International Journal of Bifurcation and Chaos}
  \textbf{\bibinfo{volume}{34}}, \bibinfo{pages}{2450129}
  (\bibinfo{year}{2024}).

\bibitem[{\citenamefont{Yan et~al.}(2013)\citenamefont{Yan, Chaudhary,
  Chul~Bae, Lewis, and Granick}}]{yan2013colloidal}
\bibinfo{author}{\bibfnamefont{J.}~\bibnamefont{Yan}},
  \bibinfo{author}{\bibfnamefont{K.}~\bibnamefont{Chaudhary}},
  \bibinfo{author}{\bibfnamefont{S.}~\bibnamefont{Chul~Bae}},
  \bibinfo{author}{\bibfnamefont{J.~A.} \bibnamefont{Lewis}}, \bibnamefont{and}
  \bibinfo{author}{\bibfnamefont{S.}~\bibnamefont{Granick}},
  \bibinfo{journal}{Nature Communications} \textbf{\bibinfo{volume}{4}},
  \bibinfo{pages}{1516} (\bibinfo{year}{2013}).

\bibitem[{\citenamefont{Lizarraga and
  de~Aguiar}(2020)}]{lizarraga2020synchronization}
\bibinfo{author}{\bibfnamefont{J.~U.} \bibnamefont{Lizarraga}}
  \bibnamefont{and} \bibinfo{author}{\bibfnamefont{M.~A.}
  \bibnamefont{de~Aguiar}}, \bibinfo{journal}{Chaos: An Interdisciplinary
  Journal of Nonlinear Science} \textbf{\bibinfo{volume}{30}},
  \bibinfo{pages}{053112} (\bibinfo{year}{2020}).

\bibitem[{\citenamefont{O'Keeffe et~al.}(2022)\citenamefont{O'Keeffe, Ceron,
  and Petersen}}]{o2022collective}
\bibinfo{author}{\bibfnamefont{K.}~\bibnamefont{O'Keeffe}},
  \bibinfo{author}{\bibfnamefont{S.}~\bibnamefont{Ceron}}, \bibnamefont{and}
  \bibinfo{author}{\bibfnamefont{K.}~\bibnamefont{Petersen}},
  \bibinfo{journal}{Physical Review E} \textbf{\bibinfo{volume}{105}},
  \bibinfo{pages}{014211} (\bibinfo{year}{2022}).

\bibitem[{\citenamefont{Yoon et~al.}(2022)\citenamefont{Yoon, O’Keeffe,
  Mendes, and Goltsev}}]{yoon2022sync}
\bibinfo{author}{\bibfnamefont{S.}~\bibnamefont{Yoon}},
  \bibinfo{author}{\bibfnamefont{K.}~\bibnamefont{O’Keeffe}},
  \bibinfo{author}{\bibfnamefont{J.}~\bibnamefont{Mendes}}, \bibnamefont{and}
  \bibinfo{author}{\bibfnamefont{A.}~\bibnamefont{Goltsev}},
  \bibinfo{journal}{Physical Review Letters} \textbf{\bibinfo{volume}{129}},
  \bibinfo{pages}{208002} (\bibinfo{year}{2022}).

\bibitem[{Note1()}]{Note1}
Note1, \bibinfo{note}{in the co-moving frame we are in $\theta = \omega t$}.

\bibitem[{\citenamefont{Abrams and Strogatz}(2004)}]{abrams2004chimera}
\bibinfo{author}{\bibfnamefont{D.~M.} \bibnamefont{Abrams}} \bibnamefont{and}
  \bibinfo{author}{\bibfnamefont{S.~H.} \bibnamefont{Strogatz}},
  \bibinfo{journal}{Physical Review Letters} \textbf{\bibinfo{volume}{93}},
  \bibinfo{pages}{174102} (\bibinfo{year}{2004}).

\bibitem[{\citenamefont{Kuramoto and
  Battogtokh}(2002)}]{kuramoto2002coexistence}
\bibinfo{author}{\bibfnamefont{Y.}~\bibnamefont{Kuramoto}} \bibnamefont{and}
  \bibinfo{author}{\bibfnamefont{D.}~\bibnamefont{Battogtokh}},
  \bibinfo{journal}{arXiv preprint cond-mat/0210694}  (\bibinfo{year}{2002}).

\bibitem[{\citenamefont{Hong et~al.}(2023)\citenamefont{Hong, O'Keeffe, Lee,
  and Park}}]{hong2023swarmalators}
\bibinfo{author}{\bibfnamefont{H.}~\bibnamefont{Hong}},
  \bibinfo{author}{\bibfnamefont{K.~P.} \bibnamefont{O'Keeffe}},
  \bibinfo{author}{\bibfnamefont{J.~S.} \bibnamefont{Lee}}, \bibnamefont{and}
  \bibinfo{author}{\bibfnamefont{H.}~\bibnamefont{Park}},
  \bibinfo{journal}{Physical Review Research} \textbf{\bibinfo{volume}{5}},
  \bibinfo{pages}{023105} (\bibinfo{year}{2023}).

\bibitem[{\citenamefont{O'Keeffe and Hong}(2022)}]{o2022swarmalators}
\bibinfo{author}{\bibfnamefont{K.}~\bibnamefont{O'Keeffe}} \bibnamefont{and}
  \bibinfo{author}{\bibfnamefont{H.}~\bibnamefont{Hong}},
  \bibinfo{journal}{Physical Review E} \textbf{\bibinfo{volume}{105}},
  \bibinfo{pages}{064208} (\bibinfo{year}{2022}).

\bibitem[{\citenamefont{Hao et~al.}(2023)\citenamefont{Hao, Zhong, and
  O'Keeffe}}]{hao2023attractive}
\bibinfo{author}{\bibfnamefont{B.}~\bibnamefont{Hao}},
  \bibinfo{author}{\bibfnamefont{M.}~\bibnamefont{Zhong}}, \bibnamefont{and}
  \bibinfo{author}{\bibfnamefont{K.}~\bibnamefont{O'Keeffe}},
  \bibinfo{journal}{Physical Review E} \textbf{\bibinfo{volume}{108}},
  \bibinfo{pages}{064214} (\bibinfo{year}{2023}).

\bibitem[{\citenamefont{O'Keeffe et~al.}(2023)\citenamefont{O'Keeffe, Sar,
  Anwar, Liz{\'a}rraga, de~Aguiar, and Ghosh}}]{o2023solvable}
\bibinfo{author}{\bibfnamefont{K.}~\bibnamefont{O'Keeffe}},
  \bibinfo{author}{\bibfnamefont{G.~K.} \bibnamefont{Sar}},
  \bibinfo{author}{\bibfnamefont{M.~S.} \bibnamefont{Anwar}},
  \bibinfo{author}{\bibfnamefont{J.~U.} \bibnamefont{Liz{\'a}rraga}},
  \bibinfo{author}{\bibfnamefont{M.~A.} \bibnamefont{de~Aguiar}},
  \bibnamefont{and} \bibinfo{author}{\bibfnamefont{D.}~\bibnamefont{Ghosh}},
  \bibinfo{journal}{arXiv preprint arXiv:2312.10178}  (\bibinfo{year}{2023}).

\bibitem[{\citenamefont{Ceron et~al.}(2023)\citenamefont{Ceron, O’Keeffe, and
  Petersen}}]{ceron2023diverse}
\bibinfo{author}{\bibfnamefont{S.}~\bibnamefont{Ceron}},
  \bibinfo{author}{\bibfnamefont{K.}~\bibnamefont{O’Keeffe}},
  \bibnamefont{and} \bibinfo{author}{\bibfnamefont{K.}~\bibnamefont{Petersen}},
  \bibinfo{journal}{Nature Communications} \textbf{\bibinfo{volume}{14}},
  \bibinfo{pages}{940} (\bibinfo{year}{2023}).

\bibitem[{\citenamefont{Sar et~al.}(2023{\natexlab{b}})\citenamefont{Sar,
  O’Keeffe, and Ghosh}}]{sar2023swarmalators}
\bibinfo{author}{\bibfnamefont{G.~K.} \bibnamefont{Sar}},
  \bibinfo{author}{\bibfnamefont{K.}~\bibnamefont{O’Keeffe}},
  \bibnamefont{and} \bibinfo{author}{\bibfnamefont{D.}~\bibnamefont{Ghosh}},
  \bibinfo{journal}{Chaos: An Interdisciplinary Journal of Nonlinear Science}
  \textbf{\bibinfo{volume}{33}}, \bibinfo{pages}{111103}
  (\bibinfo{year}{2023}{\natexlab{b}}).

\bibitem[{\citenamefont{Degond and Diez}(2023)}]{degond2023topological}
\bibinfo{author}{\bibfnamefont{P.}~\bibnamefont{Degond}} \bibnamefont{and}
  \bibinfo{author}{\bibfnamefont{A.}~\bibnamefont{Diez}},
  \bibinfo{journal}{Acta Applicandae Mathematicae}
  \textbf{\bibinfo{volume}{188}}, \bibinfo{pages}{18} (\bibinfo{year}{2023}).

\bibitem[{\citenamefont{Schilcher et~al.}(2023)\citenamefont{Schilcher, Rauter,
  and Bettstetter}}]{schilcher2023radii}
\bibinfo{author}{\bibfnamefont{U.}~\bibnamefont{Schilcher}},
  \bibinfo{author}{\bibfnamefont{C.~W.} \bibnamefont{Rauter}},
  \bibnamefont{and}
  \bibinfo{author}{\bibfnamefont{C.}~\bibnamefont{Bettstetter}}, in
  \emph{\bibinfo{booktitle}{2023 IEEE International Conference on Autonomic
  Computing and Self-Organizing Systems (ACSOS)}}
  (\bibinfo{organization}{IEEE}, \bibinfo{year}{2023}), pp.
  \bibinfo{pages}{151--156}.

\bibitem[{\citenamefont{Adorjani et~al.}(2023)\citenamefont{Adorjani, Libal,
  Reichhardt, and Reichhardt}}]{adorjani2023motility}
\bibinfo{author}{\bibfnamefont{B.}~\bibnamefont{Adorjani}},
  \bibinfo{author}{\bibfnamefont{A.}~\bibnamefont{Libal}},
  \bibinfo{author}{\bibfnamefont{C.}~\bibnamefont{Reichhardt}},
  \bibnamefont{and}
  \bibinfo{author}{\bibfnamefont{C.}~\bibnamefont{Reichhardt}},
  \bibinfo{journal}{arXiv preprint arXiv:2309.10937}  (\bibinfo{year}{2023}).

\end{thebibliography}

\end{document}